\documentclass{aa}  

\usepackage{graphicx}
\usepackage{natbib}
\usepackage{amssymb}
\usepackage{amsmath}
\usepackage{xspace}
\usepackage{color}
\usepackage[varg]{txfonts}
\newcommand{\gray}{$\gamma$-ray\xspace}
\newcommand{\grays}{$\gamma$-rays\xspace}
\newcommand{\RXJ}{RX J1713.7--3946\xspace}

\newcommand{\FGL}{2$^\text{nd}$ Fermi-LAT catalog\xspace}
\newcommand{\mkp}{}
\newcommand{\mkpn}{}
\newcommand{\mkpr}{}

\begin{document}

   \title{Analysis of GeV-band gamma-ray emission from SNR RX~J1713.7-3946}


   \author{S. Federici
          \inst{1}
          \and
          M. Pohl\inst{1,2}\fnmsep\thanks{Corresponding author, \email{marpohl@uni-potsdam.de}}
	  \and
	  I. Telezhinsky\inst{1,2}\fnmsep\thanks{Corresponding author, \email{igor.telezhinsky@desy.de}}
          \and
	  A. Wilhelm\inst{1,2}
	  \and
	  V.V. Dwarkadas
          \inst{3}}

\institute{DESY, 15738 Zeuthen, Germany 
\and Institute of Physics and Astronomy, University of Potsdam, 14476 Potsdam, Germany
\and University of Chicago, Department of Astronomy\& Astrophysics, 5640 S Ellis Ave, TAAC 55, Chicago, IL 60637, U.S.A.}

\date{Received ; accepted }

 
  \abstract
   {\RXJ is the brightest shell-type Supernova remnant (SNR) of the TeV \gray sky. Earlier Fermi-LAT results on low-energy \gray  emission suggested that, despite large uncertainties in the background determination, the spectrum is inconsistent with a hadronic origin.}
   {We update the GeV-band spectra using improved estimates for the diffuse galactic \gray emission and more than doubled data volume. We further investigate the viability of hadronic emission models for \RXJ.}
  {We produced a high-resolution map of
the diffuse Galactic \gray background corrected for the HI
self-absorption and used it in the analysis of more than 5~years
worth of Fermi-LAT data. We used hydrodynamic scaling relations and
a kinetic transport equation to {\mkp calculate} the acceleration and
propagation of cosmic-rays in SNR. {\mkp We then determined spectra of 
hadronic \gray emission from \RXJ, separately for the SNR interior and the cosmic-ray precursor region of the forward shock, and computed} flux variations
that would allow to test the model with observations. }
   {We find that \RXJ is now detected by Fermi-LAT with very high statistical significance, and the source morphology is best described by that seen in the TeV band. The measured spectrum of \RXJ is hard with index $\gamma=1.53\pm0.07$, and the integral flux above 500 MeV is $F = (5.5\pm1.1)\times10^{-9}$ photons cm$^{-2}$ s$^{-1}$. We demonstrate that {\mkp scenarios based on hadronic emission from the cosmic-ray precursor region} are acceptable for \RXJ, and we predict a secular flux increase at a few hundred GeV at the level of around 15\% over 10~years, which may be detectable with the upcoming CTA observatory.}
   {}


   \maketitle
%

\section{Introduction}
\RXJ, also known as G347.3--0.5, is a young ($t_\text{SNR}\simeq1600$ yr) shell-type SNR located in
the Galactic plane within the tail of the constellation Scorpius. This SNR was discovered in 1996
during the ROSAT X-rays all-sky survey~\citep{Pfeffermann:1996}. Its shape is slightly elliptical
with a maximum extent of $70'$, and it exhibits bright X-ray emission predominantly from the
western edge of the shell. The assumed distance $d_\text{SNR}\simeq1$ kpc implies a shell radius
$r_\text{SNR}\simeq8.7$ pc. The remnant contains an X-ray point-like source whose properties are
similar to central compact objects in other SNRs, thus suggesting \RXJ is the remnant of a core-collapse supernova.

Observations
with the Japanese ASCA satellite~\citep{Koyama:1997} showed a featureless
X-ray spectrum in the northwest shell of the remnant, clearly indicating non-thermal emission. Two
years later, new ASCA observations~\citep{Slane:1999zz}, confirmed
the absence of line emission everywhere in the remnant and showed that all parts of the
remnant had power-law with indices between $\sim2.2$ and $2.4$. More recent studies of the remnant conducted with Chandra~\citep{Uchiyama:2002mt} and
XMM-Newton~\citep{Cassam:2004} satellites clearly showed small-scale spatial variations of the photon index
ranging from 1.8 to 2.6 which are tracers of the acceleration and propagation history of recently accelerated electrons \citep{2012A&A...545A..47R}.

At higher energies two competing
radiation processes complicate the interpretation of the emission from \RXJ. In the leptonic scenario the \grays are
produced by electrons via inverse Compton scattering, and in the hadronic scenario the \grays are
due to $\pi^0$-decay from proton-proton interactions. Both scenarios can produce similar fluxes in
the GeV-TeV energy range. 

The H.E.S.S. telescope array detected \RXJ at the TeV-scale~\citep{Aharonian:2004vr} and showed a similarity between the
X-ray and TeV-band morphology. No decisive conclusion was drawn for the particle
population responsible for the emission, although the hadronic scenario was favoured.

The Large Area Telescope (LAT)~\citep{Atwood:2009ez}, the principal instrument on board
NASA's Fermi \gray satellite, measured an extended GeV \gray emission coincident with the position
of \RXJ~\citep{Abdo:2011pb}. The spectral analysis found a very hard spectral photon index
($\gamma \simeq 1.5$), which was well in agreement with a leptonic scenario. Nonetheless, this conclusion did
not exclude that protons are accelerated in this SNR.

In this paper we reanalyze Fermi-LAT data that have more than doubled in exposure since the original publication. Moreover, we develop dedicated tools to produce high-resolution background maps that take into account HI self-absorption. Our background modeling permits determining the key parameters of \gray emission from \RXJ, such as spectral index, flux and morphology, with reduced systematic uncertainty. Next, we present a model for {\mkp hadronic \gray emission from the cosmic-ray precursor of the forward shock of \RXJ,} that is in principle viable, and, what is more important, we propose a way to verify the model. Based firstly on analytic estimates and then on more detailed calculations we  predict a \gray flux increase at the level of 15\% on a time scale of 10~years. We suggest this flux variation be detectable by Cherenkov Telescope Array \citep[CTA][]{2013APh....43....3A}.

\section{Analysis and results}
\subsection{LAT data}

The LAT is a pair-conversion telescope designed to detect
photons from $\sim20$ MeV to $300$ GeV. A detailed
description of the LAT instrument is given in \citet{Atwood:2009ez}.
The data used in this study come from observations over a period of five years, from August 8, 2008
to August 13, 2013. Data analysis is performed with public analysis tools developed by the LAT
team, using the post-launch P7v6 data selection with the appropriate instrument response functions. 
To minimise the spill-over of atmospheric \grays from the Earth limb, a zenith angle cut of
$100^\circ$ is applied together with a method of correcting the exposure for the zenith cut
itself. The method chosen consists in excluding time intervals where any part of the region of
interest (ROI) is beyond the zenith-angle limit. Furthermore, data are not taken in consideration
while the observatory is transiting the South Atlantic Anomaly (SAA) or when the rocking angle exceeds
$52^\circ$.

In the analysis only photons with reconstructed energy greater than 500 MeV, for which the
68\%-containment radius of the PSF is narrower than $\sim1.5^\circ$, are used. The choice of selecting events above 500 MeV is
motivated by the earlier published spectrum and by the broad PSF at low energy. The broadening may in fact affect the analysis by leading
to systematic problems of source confusion in any densely populated region of the Galactic plane.

The data analysis is done using the publicly available Fermi-LAT \verb+ScienceTools+ version
9.31.1. In particular, spatial and spectral analyses are performed with dedicated scripts based on
the \textit{Python} likelihood tools that expand upon the command line tools provided with the Fermi Science
Tools package.

Since the analysis is performed over a large number of observed events, the binned likelihood
analysis is preferred. The region of interest (ROI) is a square region measuring
$28^\circ$ on a side and centred on $\alpha=258.39^\circ$ and $\delta=-39.76^\circ$ (J2000),
the nominal position of \RXJ. The angular binning is chosen to be $0.125^\circ$ per pixel in
stereographic projection to match the resolution of the \textit{corrected} diffuse \gray background.
To preserve accuracy in the likelihood analysis 30 energy bins are chosen, which
allows to accommodate rapid variations in the effective area with decreasing energy below
$\sim1$ GeV.

The \gray background is modelled in an acceptance region,
called the source region, that is $5^\circ$ larger than the ROI. This accounts for sources that
lie just outside the data region, but whose photons spilled into the data set.
The background model includes 174 point-like sources and 2 extended sources listed in the
\FGL~\citep{2012ApJS..199...31N}. The source 2FGL J1712.4--3941 is not considered in the model because it is
spatially coincident with \RXJ. Furthermore, we include an isotropic template\footnote{Available from the Fermi Science
Support Centre.} that accounts for extragalactic emission and residual cosmic-ray contamination. Finally we use a modified template for galactic diffuse \gray emission that shall be described in the next section.

\begin{figure}[t]
\includegraphics[width=0.49\textwidth]{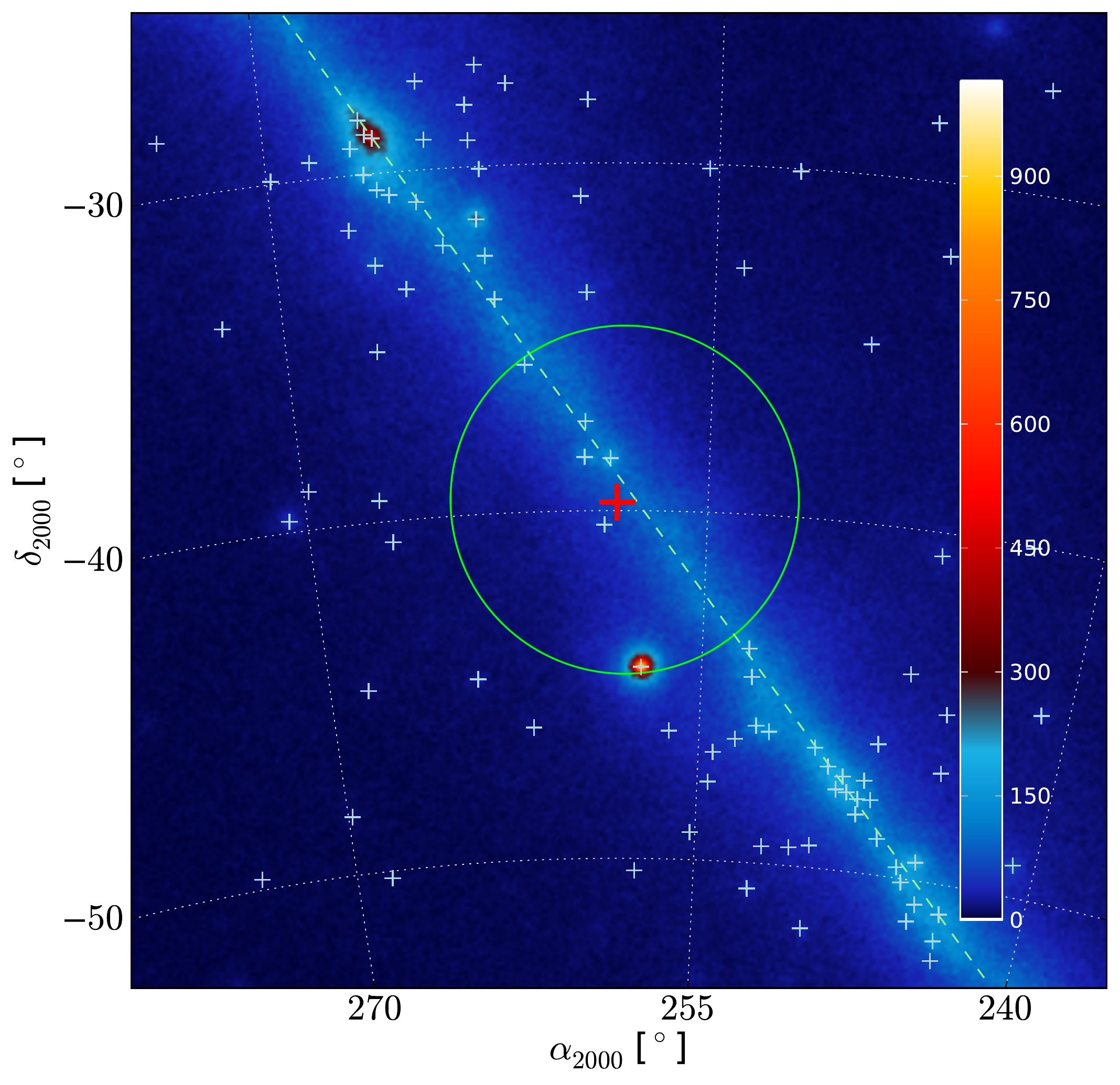}
\caption{Counts map of the region of interest (ROI) used in this work. Sources of the \FGL are
marked by light blue crosses. The green dashed line denotes the Galactic plane and the red cross
represents the nominal position of \RXJ. During the fit procedure the parameters of the sources
inside the green circular region of $5^\circ$ radius are free to vary.}
\label{fig:roi}
\end{figure}

Amongst the point-like and extended sources, 105 lie inside the ROI and most of them are modelled
with power-law energy spectra. The spatial and spectral parameters of sources located outside a circular
region of $5^\circ$ radius centred at the nominal position of \RXJ are kept fixed at the values
given in the catalog. Inside this region the spectral parameters of 7 sources along with the
normalisation of the diffuse components are permitted to freely vary. Thus, the final set of free
parameters is reduced to 20, a number which allows the fit to converge. Fig.~\ref{fig:roi} shows
the ROI with 105 sources of the \FGL and the circular region of $5^\circ$.

\subsection{The HISA contribution to the diffuse galactic \gray emission}\label{subsec:hisa_contribution}

The likelihood analysis of data from \textit{Fermi-}LAT requires templates for an isotropic component, intended to cover extragalactic emission as well as residual instrumental background, and diffuse Galactic \grays. The latter is based on infrared tracers and spectral-line information on atomic hydrogen and carbon monoxide as tracer of molecular gas.

While the absolute normalization of the templates is adjustable in the likelihood optimization, its angular structure is not. If an extended source is to be analyzed, a poor representation of the diffuse background will lead to systematic errors in the reconstructed morphology and possibly a significant offset in the derived spectrum of the source.

A major source of uncertainty is \ion{H}{i} absorption, to which a first-order 
opacity correction is often applied by
assuming a uniform spin temperature, $T_\mathrm{s}$. In terms of the brightness temperature, $T_\mathrm{b}$, and for low continuum intensity, a correction factor can be derived as
\begin{equation}\label{eq:hi_column_density}
\epsilon=-\ln\left(1-\frac{T_\mathrm{b}}{T_\mathrm{s}}\right)\,\frac{T_\mathrm{s}}{T_\mathrm{b}}\ ,
\end{equation}
where one typically uses values of $T_\mathrm{s}$ in the range 120--150~K. While this method has some merit, high-resolution \ion{H}{i} surveys have recently indicated the existence of large variations in spin temperature from one cloud to the next. Small-scale self-absorption features (HISA, as for \ion{H}{i} self-absorption) were identified that are caused by relatively cold ($\sim$ 50~K) gas clouds that are located in front of much warmer gas and sources of continuum emission. Their effect on spectral-line data goes beyond a simple correction \`a la 
Eq.~\ref{eq:hi_column_density} on account of both its absorption effect on the baseline continuum spectrum and the substantial gas mass that can be carried by the compact clouds.

We have used data from the Southern Galactic Plane Survey (SGPS)~\citep{McClureGriffiths:2005yh} to refine the standard model of galactic diffuse emission\footnote{Publicly available at the Fermi Science Support Center} to a higher intrinsic angular resolution than is provided in single-dish \ion{H}{i} surveys and to correct for HISA using the algorithm described in 
\citet{Gibson:2005xs}.

The idea behind Gibson's method
is to iteratively remove large-scale spectral and spatial emission structures from \ion{H}{i} data and to flag
the small-scale negative residuals as self-absorption features. {\mkp The additional optical depth imposed by the cold \ion{H}{i} absorber is estimated using the brightness-temperature difference between a line of sight through the absorber and other lines of sight in the vicinity, assuming most of the continuum signal is produced behind the structure. The density and mass of the absorber are calculated assuming pressure equilibrium.
Further details on the analysis can be found in Gibson's paper. }

The additional \ion{H}{i} column density derived for the absorber is then combined with that associated with the absorption correction to the original line signal. For both we calculate the location on the line of sight using the method of \citet{Pohl:2007dz}. We then use the gamma-ray emissivity derived with the GALPROP code\footnote{http://galprop.stanford.edu/},
version 54, with best-fit parameters given in \citet{Strong:2010pr}, to compute the expected \gray intensity from the additional \ion{H}{i} column density, that we found in the HISA search. Finally, we correct the standard templates of diffuse galactic \gray emission for the additional \gray signal. 

As high-resolution gas maps are available only for a narrow strip along the galactic plane, they are combined with the low-resolution maps of the Leiden-Argentina-Bonn survey \citep{Kalberla:2005ts}, and the composite map is sampled with a pixel size of
0.125$\degr$. At the location of \RXJ we thus have an accurate sampling of structure in the diffuse galactic \gray emission on scales commensurate with the angular resolution of LAT, whereas off the plane the intrinsic resolution is that of the standard galactic-emission template available at the Fermi Science Support Center.

To study the impact of the correction for the HISA on the diffuse Galactic emission (DGE) two binned likelihood analyses are
performed, one with the standard Fermi-LAT DGE template and one with the corrected DGE template. The
detailed setup of the likelihood fit, used for the standard Science Tool \emph{gtlike}, is
identical for both analyses. The two likelihood fits are performed by assuming as a spatial template
for \RXJ a uniform disk of $0.5^\circ$ radius.
The fits are performed in the 
energy range spanning from 500 MeV to 300 GeV and indicate that the revised model of galactic \gray emission in the ROI is favoured with a significance of $\sim 4\,\sigma$.

\begin{figure}
\includegraphics[width=0.49\textwidth]{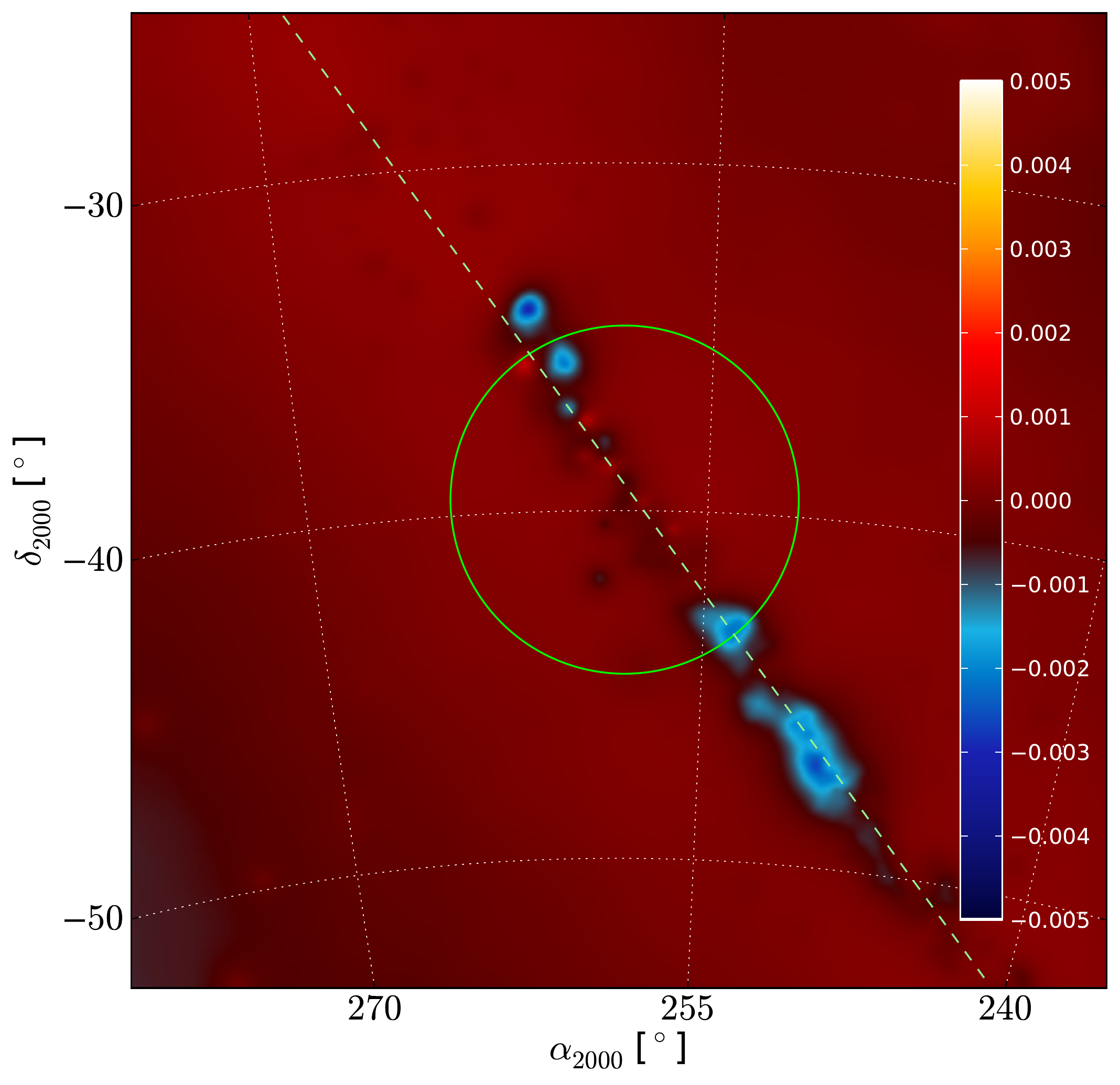}
\caption{Map of the difference in fit residuals of the revised and the standard template of galactic emission.}
\label{fig:res_models}
\end{figure}

Figure~\ref{fig:res_models} shows the difference in the best-fit model
maps of the two cases,
$[m_1-m_2]/\sqrt{m_2}$, where $m_1$ and $m_2$ refer to the standard and the revised templates, 
respectively. To be noted from the figure are the localized negative features, that correspond to HISA corrections. {\mkp At the location of \RXJ, no significant absorption feature was found, and, as we have tested, using the revised background model changes the best-fit flux from \RXJ by only about 1\%.}

\subsection{Position and spatial extension of \RXJ}\label{subsec:spatial}

\begin{figure}[t]
\includegraphics[width=0.49\textwidth]{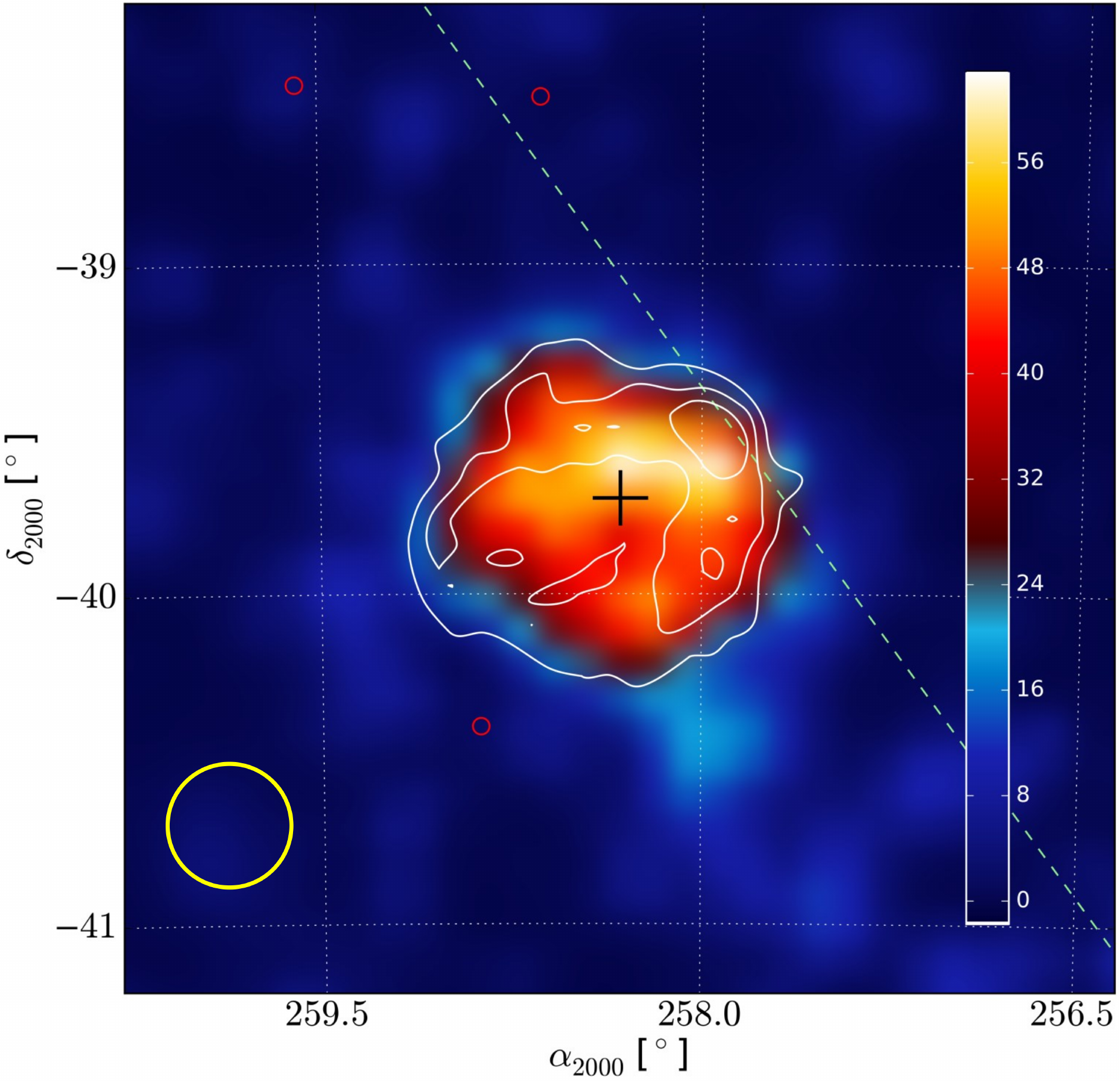}
\caption{Map of the test statistic (TS) for a point-like source in the region around \RXJ. The black
cross denotes the {\mkp best-fit centroid of a disk model} {\mkpn and the yellow circle indicates the energy-averaged 68\% confinement region of the point-spread function}. Red circles indicate the position of the second
Fermi-LAT catalog sources in the background model. Shown in white are {\mkpn contours of gamma-ray excess counts} based on H.E.S.S. observations, the levels are 25, 50, and 75.}
\label{fig:ts_map}
\end{figure}

After the first 2 years of science operation of Fermi, LAT has established a firm detection of \gray
emission from \RXJ with statistical significance of $\sim 9\sigma$. Now, 3 years later, the accumulated exposure has more than doubled, and it is expected that systematic uncertainties arising from, e.g., the spatial distribution of
emission coincident with the SNR have an impact on the reconstructed \gray spectrum. Before determining the spectrum, we therefore need to first find the localization and spatial distribution of emission from \RXJ.

To find the center of gravity of events from \RXJ, we construct a map of test statistic (TS)~\citep{Mattox:1996zz}, defined as logarithmic improvement in the likelihood function $L$,
\begin{equation}\label{eq:test_statistic}
	\text{TS} = 2\,\left(\ln L-\ln L_0\right)=2\,\ln\left(\frac{L}{L_0}\right)\ ,
\end{equation}
that we achieve by moving a putative point-like source through a grid of locations on
a square region. The region measures $\sim3^\circ$ on a side and is centered at the nominal
position of the SNR. At each point of the grid the parameters of the point-like source are free to
vary and those of all background components are fixed at the values found in the previous likelihood
analysis (see Section~\ref{subsec:hisa_contribution}). The resolution of the grid is
$0.1^\circ$ which implies 961 likelihood analyses. The resulting TS map is shown in Fig.~\ref{fig:ts_map}. {\mkp The envelope of signal with $\mathrm{TS}\ge 36$} suggests spatially extended emission, because a true point source would appear with localization uncertainty corresponding to the energy-averaged width of the point-spread function (PSF, {\mkpn indicated in Fig.~\ref{fig:ts_map} by the yellow circle}) divided by the detection significance, i.e. with $\lesssim 0.2\degr$ in diameter.
Particularly interesting is the local enhancement of the signal in the northwest region of the
shell with $\text{TS}>49$, {\mkp which matches {\mkpn well} the excess-event density seen with H.E.S.S., here indicated with white intensity contours with levels 25, 50, and 75.}

To find the best position of the source and hence minimize the systematic uncertainties, {\mkp
we then modeled \RXJ as a uniform disk with a radius of $0.5^\circ$ instead of a point-like
source. The best-fit centroid for the disk
model is indicated in Fig~\ref{fig:ts_map} and located at $\alpha=258.32^\circ$ and $\delta=-39.71^\circ$
in J2000 with an error radius of $0.02^\circ$ at the $68\%$ confidence level, which is about $0.07^\circ$ off the nominal coordinates of the source in direction of the Galactic plane.} All the following analyses are performed using spatial
templates centered at {\mkp this best-fit position}.

\begin{figure}[t]
\includegraphics[width=0.49\textwidth]{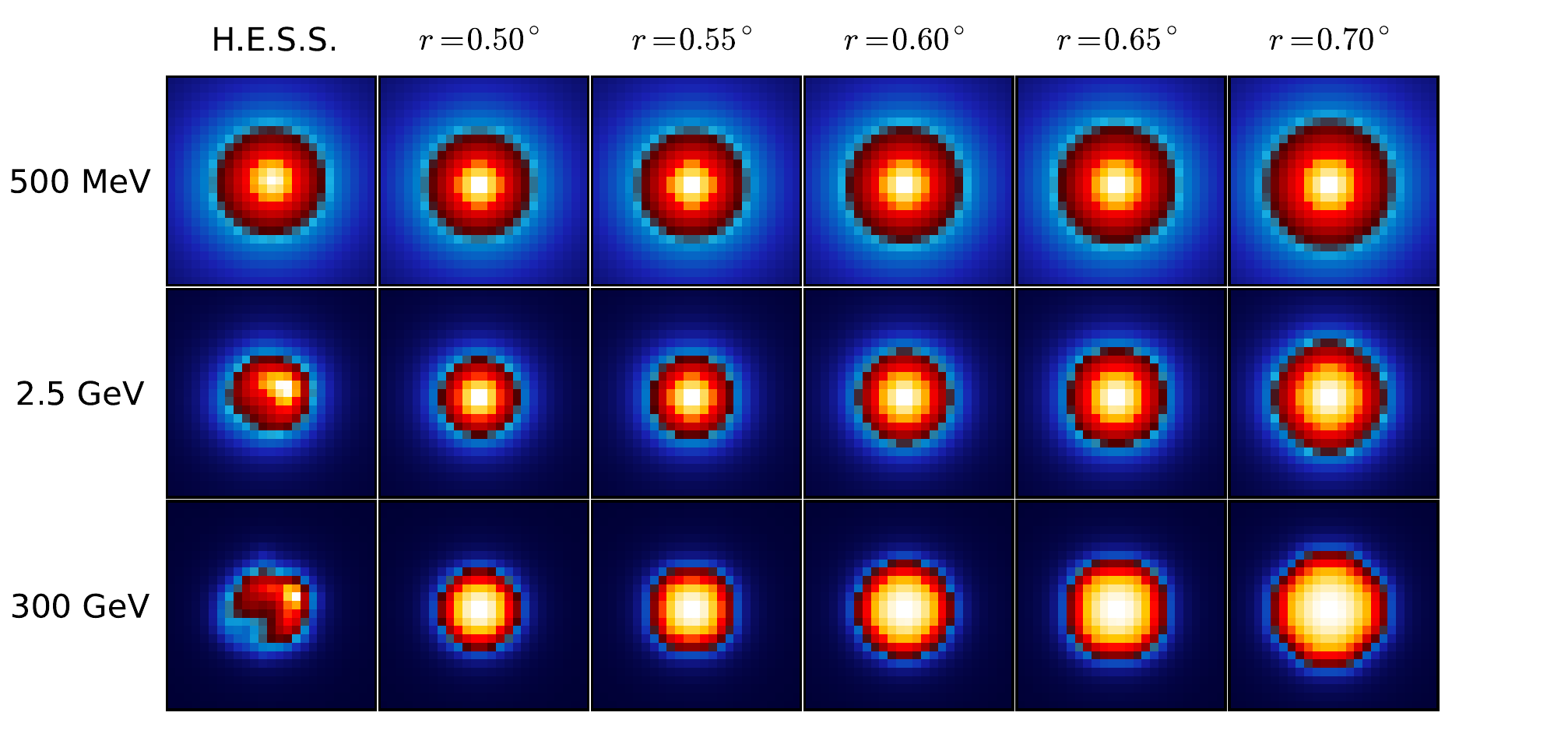}
\caption{Spatial templates used to model the emission associated with \RXJ. The templates are
convolved with the Fermi-LAT PSF and are shown for three different energies. }
\label{fig:snr_templates}
\end{figure}

To investigate the spatial morphology of the emission associated with \RXJ a number of spatial
templates are tested. Five templates are uniform disks with
radii ranging from $0.5^\circ$ to $0.7^\circ$ in steps of $0.05^\circ$.
An additional template is designed to reflect the TeV-band intensity distribution observed with the H.E.S.S. telescope. In Fig.~\ref{fig:snr_templates} we show all these spatial templates after being
convolved with the Fermi-LAT PSF. Due to the broadening
of the PSF at low energies the detailed shape and size of the six templates can only be
distinguished at high energies. Table~\ref{tbl:morphology} summarizes the best-fit parameters in the likelihood analysis of each template, performed in the energy range from 500 MeV to 300 GeV by optimizing a spectral model of the form
\begin{equation}
 \frac{dN}{dE}=N_0\,\left(\frac{E}{E_0}\right)^{-\gamma},
\end{equation}
where $N_0$ is the prefactor, $E_0$ the energy scale, and $\gamma$ is the spectral index.

All models {\mkp yield a high test statistic ($150\leq\text{TS}\leq163$), about 3 times that achieved in a point-source fit, implying that the emission
is well resolved by the LAT as an extended region. For uniform-disk templates TS slowly decreases beyond a radius of $0.55^\circ$, indicating that this value marks the extent of GeV-scale emission from \RXJ. The H.E.S.S. template fits best, and the TS value for the uniform disk of radius $0.55^\circ$ is also acceptable ($\Delta\text{TS}=2$).} {\mkpr As the H.E.S.S. template does not fit significantly better than the disk of radius $0.5^\circ$ that we used to find the centroid position, we see no reason to deviate from the sequence of analysis steps defined a priori, namely first finding the centroid position and then determining the best-fitting template. We thus use the H.E.S.S. template, centered at $\alpha=258.32^\circ$ and $\delta=-39.71^\circ$ in J2000, in the following spectral analysis. In the end, the vagaries in choosing the spatial template and centroid position contribute to the systematic-uncertainty margin that we shall discuss below.} 

\begin{table}[h]
 \caption{Morphological analysis of the \gray emission associated with
 \RXJ.}\label{tbl:morphology} 
 \centering
 \begin{tabular}{|l|c|c|c|}
 \hline
  Morphology & Flux\tablefootmark{a} & Photon index &
  TS\\
  \hline
  Disk $\ r=0.50^\circ$ & $4.982\pm0.916$ & $1.508\pm0.075$ & 160\\
  Disk $\ r=0.55^\circ$ & $5.122\pm0.941$ & $1.509\pm0.075$ & 161\\
  Disk $\ r=0.60^\circ$ & $5.449\pm1.069$ & $1.513\pm0.075$ & 159\\
  Disk $\ r=0.65^\circ$ & $5.547\pm1.052$ & $1.514\pm0.075$ & 156\\
  Disk $\ r=0.70^\circ$ & $5.908\pm1.216$ & $1.521\pm0.076$ & 150\\
  H.E.S.S.   		& $5.522\pm1.075$ & $1.528\pm0.074$ & 163\\
  \hline
 \end{tabular}
 
\tablefoottext{a}{The integral flux from \RXJ is calculated over the energy range 
500~MeV -- 300~GeV and it is given in units of $10^{-9}$ photons cm$^{-2}$ s$^{-1}$.}
\end{table}

\subsection{Energy spectrum of \RXJ}\label{subsec:spectrum}

As first step toward measuring the spectrum of \gray emission from the SNR, we perform a global likelihood analysis using the H.E.S.S. template as spatial model for \RXJ.  The fit
yields a $\gamma=1.53\pm0.07$ and the measured integral photon flux above 500 MeV is
$F = (5.52\pm1.07)\times10^{-9}$ photons cm$^{-2}$ s$^{-1}$.

In order to obtain a spectral energy distribution (SED) for the SNR, the entire energy range is
divided into 9 logarithmically spaced energy bins. For each individual bin a likelihood fit is
performed using the H.E.S.S. template as spatial model for \RXJ and the spectral model and parameters obtained in the global fit (and shown in Table~\ref{tbl:morphology}). For those energy bins
where TS$< 7$ a $95\%$ confidence-level flux upper limit is calculated.

{\mkp To estimate the amplitude of systematic uncertainties, the data are re-fitted for several variations of the default
model. First, the corrected diffuse-background model is replaced by the standard background model, yielding a 1\% variation in the derived flux.
Second, the H.E.S.S. template is substituted with reasonably fitting uniform-disk templates ({\mkpn radii $0.5^\circ$, $0.55^\circ$, $0.6^\circ$}), indicating an uncertainty of $\sim$6\% at 20 GeV increasing to 10\% at 500~MeV and 8\% at 300~GeV. For the worst-fitting template, the disk of $0.7^\circ$ radius, we find an offset of 10\% at 20 GeV.} Furthermore, uncertainties
due to instrumental effects are also taken into account. The flux measurement in fact depends on the
knowledge of the effective collecting area of LAT as a function of energy. The systematic error in
the effective area for the P7SOURCE\_V6 event class may be quoted as $10\%$ at 100 MeV, $5\%$ at 560
MeV, and again $10\%$ at 10 GeV and above~\citep{Ackermann:2012kna}. All these contributions {\mkp combined indicate overall systematic uncertainties at the level of about 11\% at 500~MeV, 12\% at 20~GeV, and 13\% at 300 GeV.}
\begin{figure}[t]
\includegraphics[width=0.49\textwidth]{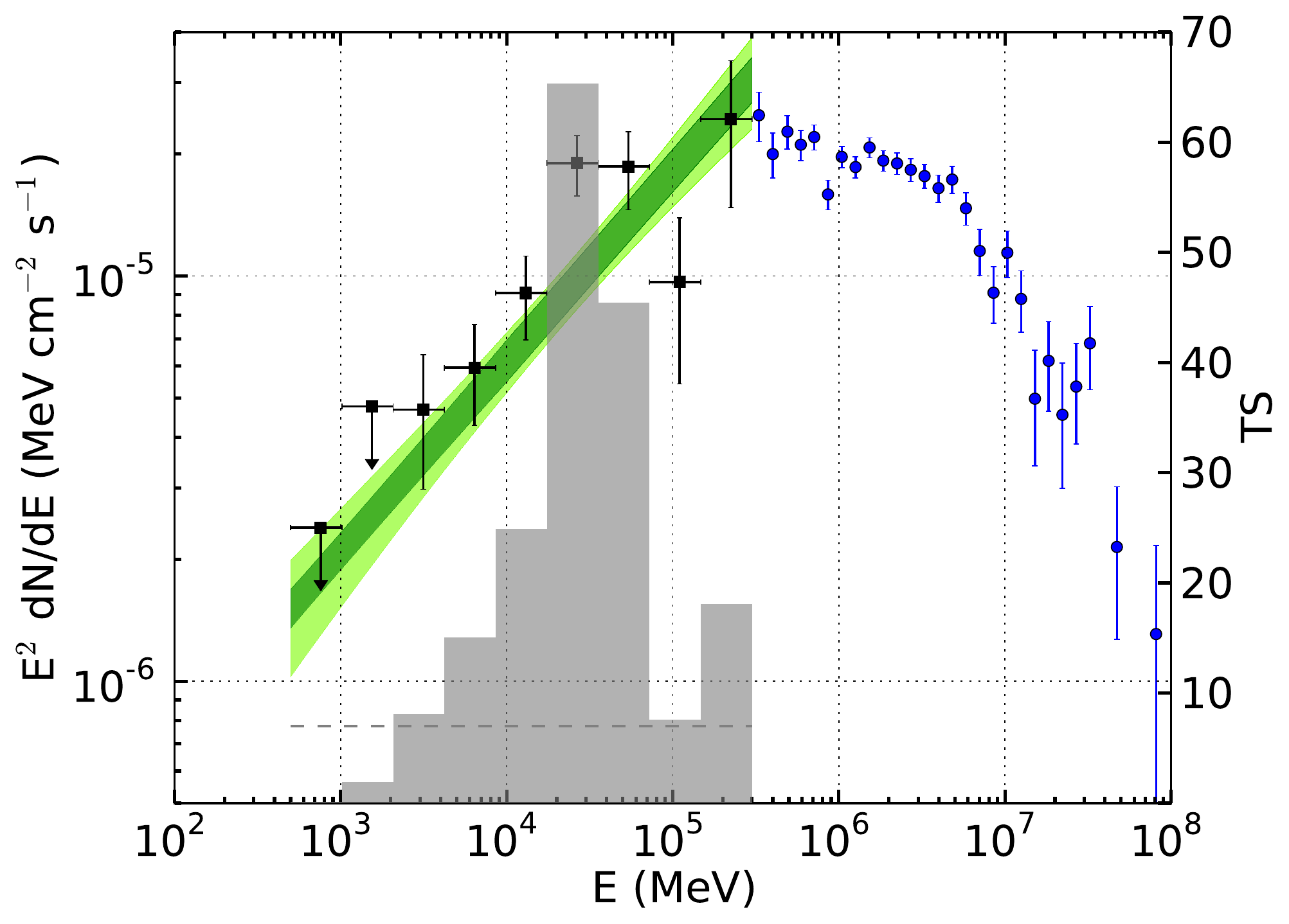}
\caption{Broadband \gray  spectrum of \RXJ as measured here (black points) and by H.E.S.S.
(blue points). The two bow-tie bands denote systematic (dark green) and {\mkp total (light green) uncertainties of the single-power-law fit, the latter assuming that systematic and statistical uncertainties can be added in quadrature.} The
histogram indicates the TS value for each energy bin. The dotted line is the threshold for setting a
$95\%$ confident level upper limit.}
\label{fig:spectrum}
\end{figure}

\begin{table}[h]
 \caption{Spectral data points of GeV-scale \gray emission associated with
 \RXJ.}\label{tbl:spectrum} 
 \centering
 \begin{tabular}{|r|r|r|}
 \hline
  $E_\mathrm{mean}$\tablefootmark{a} & $E\,F(E)$ & $\delta \left[E\,F(E)\right]$ \\
  \hline
 $7.59\cdot 10^{2}$ & $2.40$ & $0.59$ \\
 $1.55\cdot 10^{3}$ & $4.77$ & $1.58$ \\
 $3.14\cdot 10^{3}$ & $4.68$ & $1.71$ \\
 $6.40\cdot 10^{3}$ & $5.94$ & $1.66$ \\
 $1.30\cdot 10^{4}$ & $9.08$ & $2.12$ \\
 $2.65\cdot 10^{4}$ & $19.0$ & $3.23$ \\
 $5.40\cdot 10^{4}$ & $18.6$ & $4.06$ \\
 $1.10\cdot 10^{5}$ & $9.67$ & $4.25$ \\
 $2.24\cdot 10^{5}$ & $24.4$ & $9.60$ \\
  \hline
 \end{tabular}
 
\tablefoottext{a}{Energies are in MeV and the $E\,F(E)$ flux from \RXJ is calculated in units of $10^{-6}$ MeV cm$^{-2}$ s$^{-1}$.}
\end{table}

The spectral data points are given in Table~\ref{tbl:spectrum}, and the resulting energy spectrum of \RXJ is shown in Fig.~\ref{fig:spectrum} together with
previously published H.E.S.S. measurements \citep{2011A&A...531C...1A}. The two green bands indicate the systematic and the total
uncertainties in a fit of a single power law. The 9 Fermi-LAT data points (black crosses) only include
statistical errors. In the background of the figure the gray histogram represents the distribution
of TS values obtained for each bin of the Fermi-LAT SED. The dotted line at the bottom of the
histogram denotes the threshold ($\text{TS}=7$) for setting a $95\%$ confident-level upper limit on
the Fermi-LAT data points. The physical interpretation of the SED will be discussed in the next
section.

\section{Interpretation}\label{sec:discussion}

The hard spectrum of the GeV-band emission initially discovered by \citet{Abdo:2011pb} and confirmed here suggests an inverse Compton (IC) origin of the emission. In a simple one-zone scenario for the particle distribution and subsequent emission it is difficult to accommodate a strong hadronic component. In this section we point out that more complex scenarios are possible, although they require some degree of fine-tuning. In an analytic assessment and detailed calculations we explore the viability of a hadronic-emission scenario involving radiation from the vicinity of \RXJ.

{\mkp One-zone models of the broadband spectral energy distribution (SED) from X-ray~\citep{Tanaka:2008zg} to GeV \gray (Fermi-LAT), and TeV \gray (H.E.S.S.) energies invariably wash over, e.g., the variations in the X-ray spectrum across the remnant \citep{2014arXiv1401.7418S}, which may reflect electron ageing and a changing magnetic field \citep{2012A&A...545A..47R,2014ApJ...790...85R}. A few general insights can be gained, though.
Stringent constraints on the gas density in \RXJ are given by the absence of a clear detection of thermal X-ray emission~\citep{Tanaka:2008zg}, indicating a post-shock gas density of $n_\text{H}\lesssim 0.8$ cm$^{-3}$. 

Given the low gas density and the $\gamma$-ray brightness of \RXJ, hadronic models require a very large energy density in cosmic rays and a very hard particle spectrum starting from lowest energies to reproduce the observed GeV--TeV \gray spectrum. This is difficult to explain by diffusive shock acceleration even in non-linear mode. Leptonic models, on the other hand, invariably require an unusually weak magnetic field of amplitude $\sim 10\ \mathrm{\mu G}$, implying that electron energy losses are negligible and, {\mkpn at odds with theoretical expectations for efficient diffusive shock acceleration,} that there is no significant amplification of magnetic field at the shock. Also, the electron spectrum must be moderately soft ($\alpha\gtrsim 2$) to simultaneously account for X-ray, Fermi and H.E.S.S. data.
There are more complicated hadronic models, however, which may fit the GeV-scale emission from \RXJ. We shall discuss them below.}

\subsection{Hadronic emission from the SNR interior}
\label{IntHad}

\RXJ was likely produced by a core-collapse supernova. The progenitor star would have emitted a wind that pushed the ambient medium into a dense gas shell at some distance. {\mkp The gas} density inside the wind zone is low, implying a weak deceleration of the forward shock as is observed. The high velocity currently measured suggests that the forward shock has not yet reached the wall enclosing the wind-blown cavity.

\citet{2012ApJ...744...71I} suggested that dense gas clouds survive passage through the forward shock of \RXJ. Slow and energy-dependent diffusion of cosmic rays into these clouds would then lead to a hard particle spectrum at the {\mkp high-density} core of the cloud. \citet{2014arXiv1406.2322G} demonstrate that a fit to the GeV-to-TeV spectrum can be achieved, albeit apparently requiring a mass of $\sim 500\ M_{\sun}$ inside the SNR, carried only by massive clouds that GeV--TeV-scale cosmic rays cannot fully penetrate. This is much more than can condense out of the wind of the massive progenitor star of \RXJ, and so the scenario would require that pre-existing clouds have been unaffected by the wind. Besides the fact that non-penetration of clouds is not observed in the \gray spectra of nearby molecular complexes \citep{2012ApJ...755...22A}, we see two difficulties with this scenario.

Firstly, the medium around massive stars tends to be largely homogenized before they explode as core-collapse Supernova, owing to photo-evaporation and the rocket effect \citep{1984ApJ...278L.115M}. 
Secondly, as the stellar wind streams around dense gas clouds, Kelvin-Helmholtz instabilities followed by Rayleigh-Taylor instabilities will disrupt and wash off their outer layers, leading to large streams of gas with moderate density ($n\approx 1 - 10 \ \mathrm{cm^{-3}}$) that upon contact with the SNR forward shock would emit intense X-ray emission. One would have to demonstrate that its intensity is compatible with the published upper limits for thermal X radiation.

A large fraction of the gas clouds and the outer gas shell will thus be located outside of the SNR, where they would be illuminated by the cosmic-ray precursor to the forward shock. This scenario we shall discuss in the next subsection.

\subsection{Hadronic emission from the cosmic-ray precursor}
\label{ExtHad}

The width of the cosmic-ray precursor to astrophysical shocks is strongly energy dependent on account of the momentum dependence of the mean free path for scattering. At a given location, this leads to very hard particle spectra with cut off at an energy determined by the distance from the shock, the diffusion coefficient, and the shock velocity \citep{2009MNRAS.396.1629G,2011ApJ...731...87E,Teletal12b}. {\mkp It has been argued before that the escape of cosmic rays from SNRs might lead to strong TeV-band $\gamma$-ray emission from molecular clouds located 10\,pc or more from the remnant. In the case of \RXJ, the target gas has to be located very close to the forward shock, otherwise we cannot observe intense emission at 50--500~GeV. Consequently, it is illuminated not primarily by so-called runaway cosmic rays that have escaped from the system, but by the cosmic-ray precursor to the forward shock.} We do not doubt that it would be possible to find parameters that permit reproducing the observed \gray spectrum of \RXJ within such a scenario. Rather, we predict a gradual but detectable increase in \gray flux over a decade or two that renders this scenario testable. To substantiate our claim, we shall first present an analytic estimate to illustrate why the signal would grow with time. We will then use a more sophisticated model to calculate the time dependence of the \gray spectrum.

\subsubsection{Analytic estimate}
\label{sec:toy}
We shall begin with an analytic model that despite its simplicity will demonstrate the existence and magnitude of time dependence in the \gray signal from a gas cloud just outside of an SNR.
The transport equation for the differential number density of cosmic-rays, $N(E,r)$, comprises terms for spatial diffusion and advection, that assuming spherical symmetry can be written as
\begin{align}
\frac{\partial N(E,r)}{\partial t}&-\frac{1}{r^2}\,\frac{\partial}{\partial r}\,\left[r^2\,D(E,r)\,\frac{\partial N(E,r)}{\partial r}-r^2\,V(r)\,N(E,r)\right]\nonumber \\
& +{\cal D}_E\,N(E,r)=Q(E,r)\ ,
\label{eq:m1}
\end{align}
where ${\cal D}_E$ denotes the differential operator in energy.
Rewriting Eq.~\ref{eq:m1} in co-moving coordinates anchored on the shock position, $r_\mathrm{s}$, 
\begin{equation}
r\ \rightarrow\ \ \rho=r-r_\mathrm{s}\quad \mathrm{and}\quad V_\rho =V-{\dot r_\mathrm{s}}
\label{eq:m2}
\end{equation}
and restricting to the vicinity of the forward shock ($\rho\ll r_\mathrm{s}$),
we may assume a plane-parallel geometry. As particle acceleration is faster than the hydrodynamical evolution of the SNR, the particle distribution in co-moving coordinates can be in good approximation written in the steady-state limit,
\begin{equation}
-\frac{\partial}{\partial \rho}\,\left[D(E,\rho)\,\frac{\partial N(E,\rho)}{\partial \rho}-V_\rho\,N(E,\rho)\right] +{\cal D}_E\,N(E,\rho)=Q(E,\rho)\ .
\label{eq:m3}
\end{equation}
Note that in the shock precursor region sources are absent, i.e. $Q(E,\rho)=0$.
The term containing ${\cal D}_E$ is expected to be small, because energy-changing processes operating in the precursor region are very slow compared to the SNR evolution, and so we have to search for homogeneous solutions to the spatial part of the transport equation, 
\begin{equation}
\frac{\partial}{\partial \rho}\,\left[D(E,\rho)\,\frac{\partial N(E,\rho)}{\partial \rho}-V_\rho\,N(E,\rho)\right]\simeq 0\ .
\label{eq:m3a}
\end{equation}
Generally, there are 2 solutions to Eq.~\ref{eq:m3a} that are approximately
\begin{equation}
N_1\simeq \frac{A}{V_\rho}\qquad N_2\simeq C\,\exp\left(\int^{\rho} d\rho'\ \frac{V_\rho}{D(E,\rho')}\right)
\ ,
\label{eq:m4}
\end{equation}
where $A$ and $C$ are constants that are determined by matching the cosmic-ray density at the shock. The plasma flow speed in the precursor region can be somewhat modified by non-linear shock modification, and so we only know $V_\rho\gtrsim - {\dot r_\mathrm{s}}$. It is obvious that only $N_2$ satisfies the boundary condition $N(\rho=\infty)=0$, therefore it must describe the true cosmic-ray density profile reasonably well. 

Now suppose a gas cloud is located at radius $r_0$, corresponding to the time-dependent co-moving coordinate $\rho_0 (t)=r_0 - r_\mathrm{s} (t)\simeq r_0 - \dot r_\mathrm{s}\,t$, where the last expression is approximately valid for short time intervals. The gas cloud is illuminated with cosmic rays of differential density
\begin{equation}
N\simeq N_0 (E)\,\exp\left(\int^{\rho_0} d\rho'\ \frac{V_\rho}{D(E,\rho')}\right)\ .
\label{eq:m6}
\end{equation}
The time scale on which the cosmic-ray illumination changes can be estimated as 
\begin{equation}
\tau \simeq \frac{N}{\vert\frac{\partial N}{\partial t}\vert}\simeq \frac{D(E,\rho_0)}{{\vert\dot \rho_0}\, V_\rho\vert}\simeq \frac{D(E,\rho_0)}{\dot r_\mathrm{s}^2}\ .
\label{eq:m7}
\end{equation}
If we scale the diffusion coefficient to the Bohm limit, $D=\eta\, r_\mathrm{L}\,c/3$, we find for the time scale of variation
\begin{equation}
\tau \simeq (10\ \mathrm{yrs})\ \eta\,\left(\frac{E}{10\,\mathrm{TeV}}\right)
\,\left(\frac{B}{10\,\mathrm{\mu G}}\right)^{-1}\,\left(\frac{\dot r_\mathrm{s}}{3000\,\mathrm{km/s}}\right)^{-2}\ .
\label{eq:m8}
\end{equation}
Protons with 10~TeV in kinetic energy produce \grays at energies around 500~GeV, at which imaging atmospheric Cherenkov telescope have their peak sensitivity.
As long as the diffusion coefficient is not significantly larger than the Bohm limit, we expect observable variations in the \gray flux from bright, young SNRs such as \RXJ, in particular if streaming instabilities amplify the magnetic field to amplitudes in excess of $10~\mu$G.

To be noted is that the variation time scale does not explicitly depend on the distance from the shock. Implicitly, a certain distance dependence arises from the fact that the diffusion coefficient must be evaluated near the location of the gas cloud, not at the shock.

\subsubsection{Detailed modeling}

In this subsection we describe a more detailed model of \RXJ and its high-energy (HE) \gray flux variability that we expect according to the predictions outlined in subsection \ref{sec:toy}. We consider a system consisting of a core-collapse SNR and a shell of dense gas located just outside the remnant. We do not pretend to give a good broadband fit to the data because that would require a sophisticated model of the evolution of, and subsequent emission from, the remnant itself \citep[e.g.][]{Teletal13}. In any case, the remnant itself is a rather faint source of \gray emission on account of the low-density environment left by the progenitor stellar wind. The hadronic emission is dominated by \grays from the dense shell that is illuminated by high-energy cosmic-rays {\mkp in the precursor}. Consequently, the overall spectrum is expected to be hard. In low-energy (LE) \grays the flux from the SNR and the shell may be comparable, because a low gas density near the forward shock is compensated by a high intensity of LE cosmic rays. The flux of HE \grays from the dense shell and its variation depend on the cosmic-ray diffusion coefficient and the current kinematics of the shock, parameters that change little if sophisticated models of SNR are considered. Therefore, the prediction of HE \gray flux variation offered here is robust.

We numerically solve the full transport equation for cosmic rays (\ref{eq:m1}) on a grid that extends from the SNR center to several dozens of SNR radii ahead of the forward shock (FS). The grid extension allows to account for {\mkp the escape from the precursor of} particles of both species, protons and electrons. We can use either full hydrodynamical simulations or analytic solutions to trace the evolution of the SNR, i.e. the velocity and radius of shocks and the plasma flow profile inside the remnant. A crucial parameter for particle acceleration is the diffusion coefficient. We assume it to be Bohm-like, $D_{B} = \eta\, r_{L}\, v/3 $, where $\eta = 1.0$ (at the shock), $r_{L}$ is the particle Larmor radius, and $v$ is the particle velocity. Ahead of the shock the diffusion coefficient linearly increases \citep{Bel78, Dru83} until thirty SNR radii ahead of the shock it reaches  $D_{G} = 10^{29} \left({E}/ {10\textrm{ GeV}}\right)^{1/3} \left({B}/ {3\mu\textrm{G}}\right)^{-1/3}$, a value typical for the galactic propagation. {\mkp Close to the shock diffusion is still Bohm-like, and so a typical cosmic-ray precursor will be established. Further ahead, the probability of escape from the system becomes significant, and particles may leave the system.} For full details we refer a reader to a series of publications \citep{Teletal12a, Teletal12b, Teletal13}.

Given that thermal X-rays are not detected, the current number density of gas upstream of the blast wave must be around 0.05~cm$^{-3}$, much lower than the typical density in the interstellar medium. Therefore, the explosion should have taken place in the wind-blown bubble of the massive progenitor star, and the SN may be classified as a core-collapse event. To reconstruct these parameters we used analytical solutions for the free-expansion stage of SNR expanding into a wind-blown cavity \citep{TruMcK99}. The current SNR radius $R \simeq 8.65$~pc and shock speed $V_{sh} \simeq 4200$~km/s (for a distance of 1~kpc) permit a range of parameter values for the mass-loss rate and the progenitor wind speed. As example we here assumed for the mass-loss rate $\dot M \simeq 10^{-4} M_{\sun}$, for the wind speed $v_\mathrm{w} \simeq 50$~km/s, and for the explosion energy $E_{SN} = 10^{51}$~ergs, corresponding to a relatively fast red-supergiant wind. It serves as demonstration of the stability of our results against the choice of hydrodynamical parameters that we even tested a Sedov solution, and the implications for the flux increase at a few 100~GeV were virtually the same. The magnetic field in the precursor region is assumed to be $B_{p} = 23 \mu$G which is shock-compressed to a downstream field $B_{d} = 75 \mu$G inside the SNR \citep{Bal06, Aceetal09}.

{\mkp \RXJ is considered to be 1600~years old, and we use that as fiducial age of \RXJ. Fitting the Truelove-McKee profiles permits a certain liberty in choosing the distance, the age, and the current shock velocity. Note that the actual age is of minor relevance for the scenario at hand, important are the time span between 2 observations of the object and, as Eq.~\ref{eq:m7} illustrates, the shock velocity and the diffusion coefficient. The allowed range in current shock velocity is fairly small, thus setting limits on the age, and despite entering Eq.~\ref{eq:m7} quadratically it does not lead to a large uncertainty in the timescale of flux variability. The location of the dense gas shell is prescribed by the necessity to reproduce the presently observed \gray spectrum from \RXJ, as the precursor of cosmic rays at 10~TeV should just reach it. That leaves the diffusion coefficient and in particular the efficacy of magnetic-field amplification as the main sources of uncertainty, for which we use conservative assumptions.}

\begin{figure}[!t]
\includegraphics[width=0.49\textwidth]{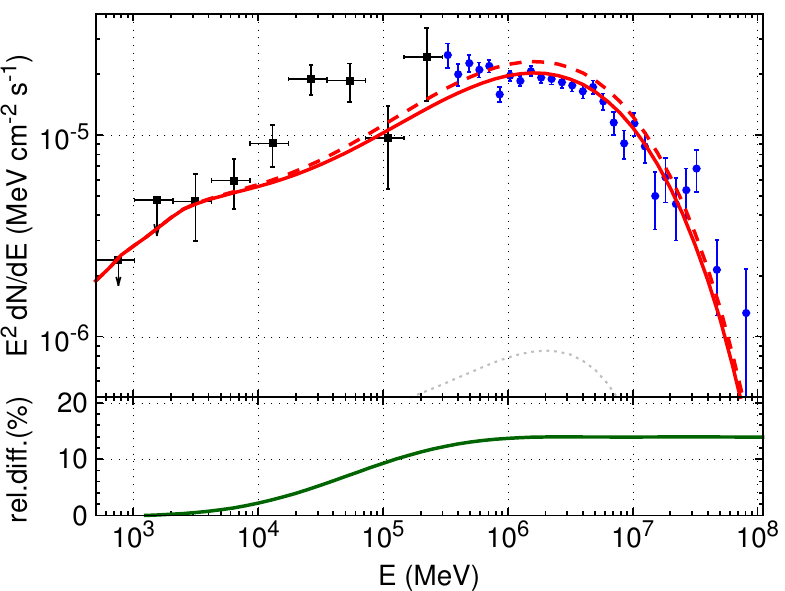}
\caption{Broadband \gray  spectrum of \RXJ as measured with LAT (our results, black points) and with H.E.S.S. (blue points), compared to calculated HE \gray spectra at the age of 1600~years (solid red line) and 1610~years (dashed red line). The IC contribution to both spectra is indicated for additional information (gray dotted line). The relative difference between spectra at two epochs is shown in the lower panel (solid green line).}
\label{fig:rad_dif_map}
\end{figure}

We numerically solve the time-dependent cosmic-ray transport equation (Eq.~\ref{eq:m1}) using the above parameters \citep{Teletal12a}. We then calculate emission spectra and intensity maps  at two epochs, at 1600 and at 1610~years, including pion-decay \citep{Huang:2006bp}, inverse Compton (IC) and synchrotron emission in turbulent magnetic field \citep{2014arXiv1411.2891P}. The dense gas shell radially extends from $8.75$~pc to $9.05$~pc from the SNR center, and so the shock does not interact with the dense material, but CRs in the shock precursor can efficiently reach the dense gas. A small separation between the gas shell and the forward shock is necessary to reproduce the $\nu\,F_\nu$-peak at $\sim$1~TeV in the hadronic \gray emission. The results of our calculations are presented at Fig.~\ref{fig:rad_dif_map} (solid red line). The HE \gray spectrum at the age of 1600~years is normalized to fit H.E.S.S. data and to reconstruct the measured flux, $F = (17.2 \pm 0.4) \times 10^{-12}$~photons~cm$^{-2}$~s$^{-1}$ \citep{2011A&A...531C...1A}. 

The parameters are defined by the gamma-ray flux ratio between the TeV band and the GeV band as well as the location of the peak in the $\nu F_\nu$ spectrum. The latter determines the separation between the forward shock and the gas shell, somewhat depending on the scaling of the diffusion coefficient in the precursor. The former prescribes the mass or gas density in the shell, $\sim 300\ M_\odot$ and $n_\mathrm{H} \simeq 35\ \mathrm{cm}^{-3}$ for our SNR expansion model with $17\ M_\odot$ of swept-up gas in the remnant. This is simply the amount of target material needed to fit the observed gamma-ray spectrum; a thorough study of the hydrodynamical circumstances that could give rise to a shell with the required mass is beyond the scope of this paper. The IC contribution is constrained by the observed flux of synchrotron X-rays and our choice of magnetic-field strength. {\mkp Reproducing the spectral width of the \gray signal in the $\nu F_\nu$ spectrum shown in Fig.~\ref{fig:spectrum} requires a diffusion coefficient well in excess of the Bohm limit. In other words, the parameter $\eta\approx 10$ in Eq.~\ref{eq:m8}, increasing with distance from the shock, and so with $B=23\ \mu\mathrm{G}$ and $\dot r_\mathrm{s}=4200\,\mathrm{km/s}$ we expect as time scale of variation $\tau\simeq 100$~years. The numerical treatment indicates a flux increase of 15\% over ten years, consistent with the analytical estimate.}

The main prediction of our model thus is a secular increase with time of the HE \gray flux that is largely independent of the choice of parameters, provided they permit reproducing the peak in the $\nu F_\nu$ spectrum at $\sim 1$~TeV for the current size and expansion rate of \RXJ. Fig.~\ref{fig:rad_dif_map} (dashed red line) shows the evolution of the \gray over 10~years. The relative increase is about 15\% at around 500~GeV (see the green line at Fig.~\ref{fig:rad_dif_map}). Note that the IC flux will not change during this period because the amount of target photons interacting with electrons in the shock precursor does not change noticeably. {\mkp Also note that the change in morphology will be too small to be detectable with CTA.} 

The calculated flux in the 1 -- 100~TeV energy band at the age of 1600~years amounts to $F_{1600} = 17.2 \times 10^{-12}$~photons~cm$^{-2}$~s$^{-1}$, compatible with that measured with H.E.S.S., whereas 10~years later the flux would be $F_{1610} = 19.9 \times 10^{-12}$~photons~cm$^{-2}$~s$^{-1}$, with most of the increase coming from the shell. This change in the flux could be potentially detected with the CTA \citep{2013APh....43....3A}. {\mkp The systematic uncertainties in the flux scale for the original measurement with H.E.S.S. are probably too large for that, and so one would need CTA data taken at least 10 years apart.} 

\section{Conclusions}
In this work we analyzed more than 5~years of Fermi-LAT observations of the SNR \RXJ, using high-resolution \gray background maps that account for the HI self-absorption. \RXJ appears as circular source with radius $r\simeq 0.55^\circ$ that is detected at $13\sigma$ level in the energy range between 500~MeV and 300~GeV. We see a tendency in the GeV band for the same intensity enhancement in the northwest region of the remnant that is seen with H.E.S.S. and probably due to particles interacting with denser gas located there. The measured spectrum of \RXJ indeed shows a hard spectral index of $\gamma=1.53\pm0.07$ with an integral flux above 500 MeV of $F = (5.5\pm1.1)\times10^{-9}$ photons cm$^{-2}$ s$^{-1}$.
 
Phenomenological fits of the broadband SED demonstrate that in simple one-zone models for particle acceleration and emission it is hard to accomodate any pion-decay scenario. These fits also show that a common particle power-law spectrum with an exponential cutoff for electrons and protons faces some difficulty in simultaneously accounting for the X-ray, GeV \gray, and TeV \gray observations. A more complex model with additional parameters and more flexible assumptions could modify the shape of the broadband spectrum and give a better description of the \gray data. 

We discussed alternative hadronic scenarios and pointed out some
difficulties with them. In particular, we introduced a model {\mkp of
hadronic emission from \RXJ} that reproduces the hard GeV spectrum with
index $\gamma \simeq 1.5$ as emission from a shell of dense gas that
is located a short distance upstream of the forward shock. The
existence of such a shell may be related to the winds of the
progenitor star. Although this hadronic scenario is somewhat
speculative, it has the benefit of being testable: We predict the
hadronic \gray flux to increase at the level of 15\% over 10~years
which should be observable with the future CTA facility {\mkp that is projected to operate for more than 10 years}
\citep{2013APh....43....3A}. {\mkp We have verified that the growth in \gray flux is largely independent of the choice of parameters, provided they reproduce a $\nu\,F_\nu$-peak at $\sim$1~TeV. Whereas our model assumes spherical symmetry, real SNR are not that simple, and conditions are not the same in different regions. The stability of our result suggests that such a variation of parameters does not significantly reduce the secular increase in \gray flux.}

A second implication of the scenario is
that, within the next 50 years, the forward shock of \RXJ should hit
the gas shell, which would then light up in X rays and permit
unprecedented studies of shock-cloud interactions. We stress that similar intensity fluctuations 
are expected in models invoking the forward shock moving through a medium in which most of the mass is organized in dense gas clouds.
   
\begin{acknowledgements}
      Part of this work was supported by the Helmholtz Alliance for Astroparticle Physics HAP
funded by the Initiative and Networking Fund of the Helmholtz Association. VVD's research on the  high energy emission from young SNRs is supported by NASA Fermi grant NNX12A057G.
\end{acknowledgements}


\bibliographystyle{aa}
\bibliography{References}

\end{document}